# A CNN-Based Blind Denoising Method for Endoscopic Images

Shaofeng Zou[1], Mingzhu Long[1], Xuyang Wang[1], Xiang Xie[1,2], Guolin Li[3], Zhihua Wang[1]
[1]Institute of Microelectronics, Tsinghua University, Beijing, China,
[2]Graduate School at Shenzhen, Tsinghua University, Shenzhen, China,
[3]Department of Electronic Engineering, Tsinghua University, Beijing, China,
Email: xiexiang@tsinghua.edu.cn

*Abstract*—The quality of images captured by wireless capsule endoscopy (WCE) is key for doctors to diagnose diseases of gastrointestinal (GI) tract. However, there exist many low-quality endoscopic images due to the limited illumination and complex environment in GI tract. After an enhancement process, the severe noise become an unacceptable problem. The noise varies with different cameras, GI tract environments and image enhancement. And the noise model is hard to be obtained. This paper proposes a convolutional blind denoising network for endoscopic images. We apply Deep Image Prior (DIP) method to reconstruct a clean image iteratively using a noisy image without a specific noise model and ground truth. Then we design a blind image quality assessment network based on MobileNet to estimate the quality of the reconstructed images. The estimated quality is used to stop the iterative operation in DIP method. The number of iterations is reduced about 36% by using transfer learning in our DIP process. Experimental results on endoscopic images and real-world noisy images demonstrate the superiority of our proposed method over the state-of-the-art methods in terms of visual quality and quantitative metrics.

*Keywords—Wireless capsule endoscopy, blind denoising, image quality assessment, convolutional neural network*

## I. Introduction

Wireless capsule endoscopy (WCE) offers a painless and noninvasive way to identify diseases in the gastrointestinal (GI) tract. The quality of endoscopic images directly affects doctor to diagnose diseases. Due to the limited illumination of WCE system and convolution and waving natures of GI tract, the captured images often have dark areas and suffer from low contrast and severe detail loss. Therefore, image enhancement is necessary to improve the quality of endoscopic images. However, as shown in Fig. 1, the noise is also significantly amplified after image enhancement.

The noise of image captured by WCE system varies with different GI tract environments, different cameras and camera parameter settings (such as ISO, shutter speed, and aperture, etc.). The noise model also changes after non-linear enhancement or different enhancement approaches. Moreover, the ground truth of endoscopic images is hard to be obtained in practice, because it is impossible to fix the capsule at a certain position or use different ISOs to take multiple images in the same scene due to the passive random movement of the capsule in GI tracts. Thus, it is difficult to make real-world noisy endoscopic image datasets using approaches like [1] [2]. A blind denoising method without the noise model and the ground truth maybe more suitable for this situation.

Blind denoising of real noise images includes traditional denoising methods and CNN-based methods, and is generally more challenging than removing additive white Gaussian noise. Noise Clinic (NC) [3] is a blind image denoising algorithm that first estimates a Signal and Frequency Dependent noise model and then denoises the image by a multiscale adaptation of the Non-local Bayes denoising method. Neat Image (NI) [4] is a software published by ABSoft, and it can remove the noise that appearing in low light photos taken with high ISO settings. Our experiments show that they are still limited for removing noise from the enhanced images. On the other hand, so far, only a very little work has been done to develop CNN-based model for the blind image denoising. Ulyanov D et al. [5] proposed the Deep Image Prior (DIP) method using a randomly initialized neural network and the white gaussian noise as its input, which can denoise real images with thousands of iteration operations. However, this method doesn't know when to output a clean image during the iteration process, i.e., the iteration process has to be terminated manually. Besides, the thousands of iterations for a clean image reconstruction are time consumed.

In this paper, we tackle these issues by improving DIP method to denoise images without specific noise model and ground truth. In order to solve the problem that DIP method cannot know when to terminate the iteration process, a blind image quality assessment method based on MobileNet [6] is used to estimate the quality of the reconstructed images. Through transfer learning, we also reduce the number of iterations of DIP while maintaining the denoising effect.

The paper is organized as follows. Section II presents the proposed method. Section III describes the experimental results, and the paper is concluded in Section IV.

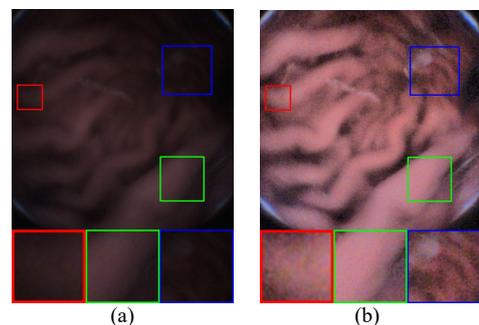

Fig. 1. Endoscopic image with dark areas and corresponding enhanced image by the method of [7]

## II. Methods

### A. Overall Framework

The architecture of the proposed method is shown in Fig. 2. Our method consists of a Blind Denoising Network (BDN)

This work was supported by Science and Technology Planning Project of Guangdong Province(2017B010110006)).

and a Blind Image Quality Assessment Network (BIQAN). The BDN is based on DIP method. The white Gaussian noise and uniform noise are the input of BDN referred from [5]. The reference images of BDN are the dark endoscopic images enhanced by [7] with obvious noise. In order to reduce the iteration number of image reconstruction, a pre-trained model is introduced for the BDN. The reconstructed images will be generated one at each iteration. Considering that the BDN cannot terminate the iteration process automatically, the BIQAN is applied to estimate the optimum quality of the reconstructed images to decide when to terminate BDN's iteration process. In BIQAN, the basic classification structure is adopted. The classification frameworks can outperform regression frameworks for ordered classes such as quality estimation. Since the input of BDN is Gaussian white noise and uniform noise, the quality of reconstructed images has obvious local fluctuations. In order to avoid falling into the local optimum and find the best quality peak point, the result of the image quality scores estimated by BIQAN are filtered by a low-pass filter for smoothing. The maximum score is found to get the best denoised image.

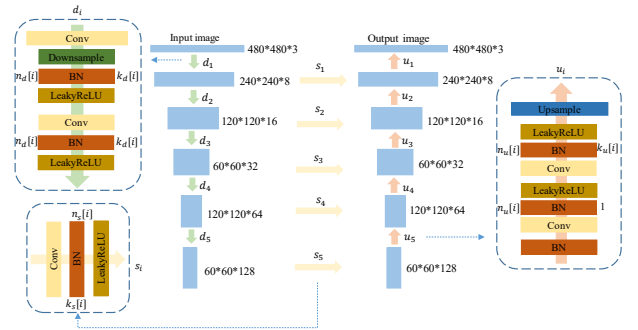

Fig. 3. The architecture of blind denoising network. $n_u[i]$, $n_d[i]$, $n_s[i]$ correspond to the number of filters at depth i for the upsampling, downsampling and skip-connections respectively. The values $k_u[i]$, $k_d[i]$, $k_s[i]$ correspond to the respective kernel sizes.

the information of the denoised image. In original DIP method, a random initial network is used to fit the noisy image. Thus, we implement the transfer learning by reconstructing the noisy image directly with the weight of network trained on a typical noisy image. In this way, the number of iterations required to achieve the best image will be reduced a lot.

The other problem is that the number of iterations required to reach the optimal reconstruction is random for each image. In the case of denoising the enhanced dark endoscopic image, the corresponding ground truth is not known, thus PSNR cannot be used to measure the quality of reconstructed image during the iteration. The best reconstructed image has to be selected manually, which is not practical. Therefore, we design a blind image quality assessment network (BIAQN) to evaluate the image quality and select the best reconstructed image without ground truth.

### C. Design of Blind Image Quality Assessment Network

Considering the excellent performance of the method proposed in [8], our approach predicts the distribution of reconstructed image quality score as a histogram instead of classifying images to low/high score or regressing to the mean score. Our proposed BIQAN stands on image classifier architecture based on MobileNet [6] due to its smaller and faster CNN architecture. MobileNet uses depthwise separable convolutions to build light weight deep neural networks. As shown in Fig. 4, we replace the last layer of the baseline CNN with a fully-connected layer with 10 neurons followed by soft-max activations.

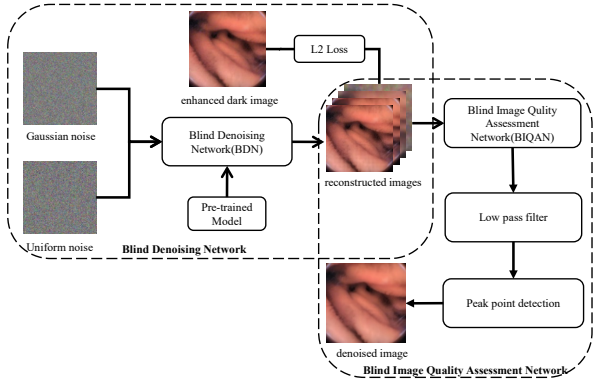

Fig. 2. Architecture of proposed method

### B. Blinding Denoising Using Deep Image Prior

In Ref. [5], Ulyanov et al. proposed Deep Image Prior (DIP) method. They show that a randomly-initialized neural network can be used as a handcrafted prior with excellent results in standard inverse problems such as denoising, super-resolution, and inpainting. As illustrated in Fig. 3, our blinding denoising network is based on DIP method using the encoder-decoder structure with skip-connections. Uniform noise is the input of network and white Gaussian noise is added as perturbation during iteration. The reference image of network is the noisy image and least square errors (L2) is used as loss function. The denoised image can be reconstructed through multiple rounds of iteration. Different from the traditional deep learning method, it can achieve excellent denoising effect without ground truth and specific noise model.

However, there are still some problems that need to be solved. The first problem is that thousands of iterations are required to achieve an optimal reconstructed image for one noisy image. It takes nearly 25 minutes to reconstruct one noisy image by iterating 5000 times on Nvidia TITAN Xp GPU. Therefore, the process of denoising is very time-consuming. Inspired by transfer learning, the network weights after multiple rounds of iterations serve as a parameterization of the restored image and already include

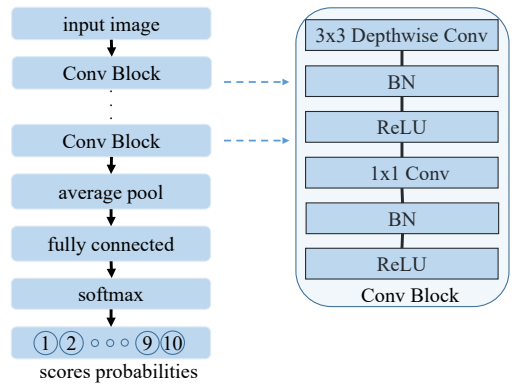

Fig. 4. Architecture of MobileNet used in our method

Most blind image quality assessment methods are mostly developed based on the commonly used datasets such as TID dataset and AVA dataset, whose images have very different properties from the reconstructed images. In order to train our proposed BIQAN, we use the reconstructed images from real-world noisy dataset with ground truth as the input of network. The PSNR of corresponding reconstructed images is converted to distribution of histogram as the metric of reconstructed image quality.

The quality rating distribution can be expressed as the empirical probability mass function $p = [p_{s_1}, p_{s_2}, ..., p_{s_n}]$, where $s_1 < s_i < s_n$, $s_i$ represents the score of i-th interval, N represents the total score interval and $p_{s_i}$ indicates the probability that the quality score falling into the i-th interval. Here, we set N=10, $s_1$=0.5, and $s_{10}$=9.5. Therefore, in order to convert PNSR to quality rating distribution, PSNR of each reconstructed image is linearly transformed to μ as follow:

$$\mu = \frac{PSNR - \min(PSNR)}{\max(PSNR) - \min(PSNR)} * 5 + 2.5 \quad (1)$$

where min (PSNR) and max (PSNR) represent the minimum and maximum PSNR from all reconstructed images respectively. Then we establish a normal distribution with mean value μ, variance σ=1.5 and number of sampling points M. Finally, we count the number of sampling points fall into i -th score interval and then calculate the corresponding distribution probability. The mean value of the distribution probability denotes as $\hat{\mu} = \sum_{i=1}^{N} s_i * p_{s_i}$, when M → ∞, $\hat{\mu} = \mu$. After the training is completed, the network can be used to evaluate the rating distribution of the reconstructed images. The mean quality score is defined as $\mu = \sum_{i=1}^{N} s_i * p_{s_i}$ to represent the quality of image in the reconstruction process.

We use Earth Mover's Distance (EMD) as our loss function, because [9] has shown that EMD-based losses can be used to train on datasets with intrinsic ordering for classification network. The EMD loss function penalizes mis-classifications according to class distances and can be expressed as:

$$EMD(p, \hat{p}) = \left(\frac{1}{N}\sum_{k=1}^{N}|CDF_p(k) - CDF_{\hat{p}}(k)|^2\right)^{1/2} \quad (1)$$

where p and $\hat{p}$ are the ground-truth and estimated probability mass functions respectively and $CDF_p(k)$ is the cumulative distribution function as $\sum_{i=1}^{k} p_{s_i}$.

The input of BDN is the Gaussian white noise, which results in the output jitters of both BDN and BIAQN. Fortunately, the change trends of their output curves are bell-shaped. To avoid falling into the local optimum, firstly, we design a simple real-time low-pass filter to smooth the quality scores of BIQAN, and then the global optimum peak can be located coarsely. Lastly, we can find the global optimum peak by exhaustive search only nearby the coarse location.

## III. EXPERIMENTAL RESULTS

In this section, we evaluate our proposed method on endoscopic image and Nam [1] dataset and compare with state-of-the-art denoising methods, including Noise Clinic(NC) [3], Neat Image(NI) [4], CBM3D [10] and CC [1].

### A. Data Preparation

Both enhanced dark endoscopic images and real-world noisy image datasets are used in our experiment. Because there are no ground truth of the endoscopic images, enhanced dark endoscopic images are only used to evaluate the visual quality of denoising. As PolyU [2] dataset has more images than Nam [1] dataset, PolyU dataset is used for training the blind image quality assessment network and Nam dataset is used for evaluating quantitative metrics. Since the size of the image in Nam dataset and PolyU dataset is particularly large, 15 regions in Nam dataset and 100 regions in PolyU dataset are cropped to size of 512 ×512 as the dataset. For training data, we first get reconstructed images and corresponding PNSR for each cropped image from PolyU dataset through BDN. Testing data is made in the same way using cropped images in Nam dataset.

### B. Training for Image Quality assessment network

For training the BIQAN, we first initialize the MobileNet using the weights pre-trained on ImageNet. The weights of the fully-connected layers are initialized randomly. During first stage of training, we freeze the convolution block and only train the fully-connected layers. A total of 10 epochs are trained using Adam with learning rate=0.001. After that, in the second stage, we still freeze the underlying convolution layer and choose to unfreeze the final convolutional layer instead of the entire network. This prevents overfitting because the entire network has a large entropy capacity and a high tendency to overfit. A total of 20 epochs are trained, still using Adam with learning rate=0.001 in this stage. The input images are rescaled to $256 \times 256$. Since the dataset is relatively small, we crop the images randomly to $224 \times 224$ and flip horizontally randomly to avoid the problem of overfitting.

### C. Performance Evaluation

We compare the proposed method with state-of-the-art denoising methods, including Noise Clinic (NC) [3], Neat Image (NI) [4], CBM3D [10]. NC and NI are two state-of-the-art blind denoising approaches, and their executable programs are publicly available. Besides, CBM3D is a state-of-the-art non-blind Gaussian denoising methods. For most non-blind denoising algorithms, the standard deviation of noise should be a parameter, so we exploit [11] to estimate the noise level of noisy image for CBM3D.

For the denoise evaluation of the enhanced dark endoscopic images, we only compare the subjective visual quality because their ground truths are unavailable. Fig. 5 shows the denoising results of an endoscopic image with different methods. Fig. 5(a) is the original image suffering from low contrast and severe detail loss. Fig. 5(b) is the enhanced image using method of [7]. For CBM3D in Fig. 5(c), it has little denoising effect due to the noise level estimated by [11] which mainly designed for Gaussian noise is too low. In red region, NI in Fig. 5(e) has less noise than NC in Fig. 5(d) but loses more structures than our proposed method in Fig. 5(f). In green region, our proposed method in Fig. 5(f) has less noise especially in dark area then NC and NI. As for blue region, our proposed method in Fig. 5(f) even denoises some pure black pixels generated in enhanced image which do not exist in original image. In general, our proposed method perform favorably in removing most noise without smoothing out image structures in comparison.

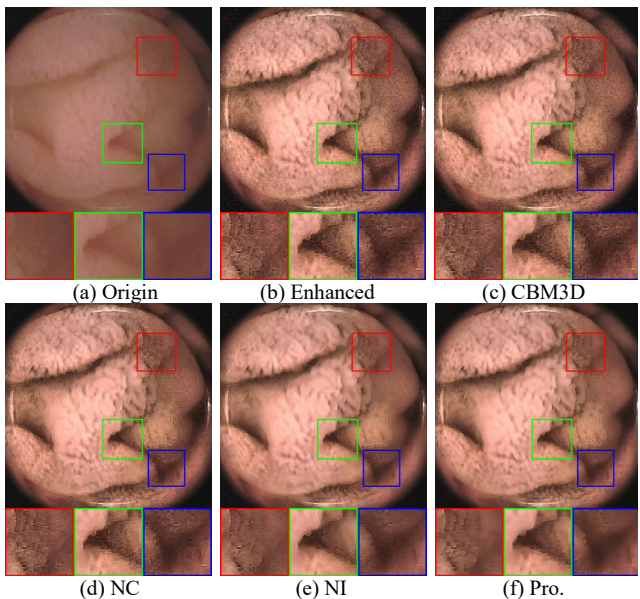

Fig. 5. Denoising results of an endoscopic image by different methods

In order to compare the objective quality fairly, we evaluate the proposed method on Nam datasets compared with NC [3], NI [4], CBM3D [10] and CC [1]. The images were taken by three different cameras and different ISOs. The PSNR results of the comparison are listed in Table I. The results of CBM3D and CC are directly copied from [1]. The highest PSNR results are highlighted in bold. On 14 out of the 15 images, our proposed method achieves the highest PSNR values. In terms of the average PSNR, our proposed method outperforms other methods 0.68dB. At the same time, CC method achieves better PSNR results than NC, NI and CBM3D. It should be noted that in the CC method, a specific model is trained for each camera and camera setting, while our proposed method denoises image without specific noise model. Results show that our method can also denoise general images effectively, which strongly increases the wide applicability of our work.

Inspired by transfer learning, we reconstruct the noisy image directly using the weight of network trained by a typical image. We evaluate our optimization on Nam dataset and choose image with serial number 12 in the Nam dataset as the typical image. As shown in Fig. 6, the number of iterations required to reach the highest PSNR shown in Fig. 6(b) is reduced by 36% while the reduction of PSNR can be barely negligible as shown in Fig. 6(a). Based on this optimization, more than one-third of time of reconstructing a clean image can be saved.

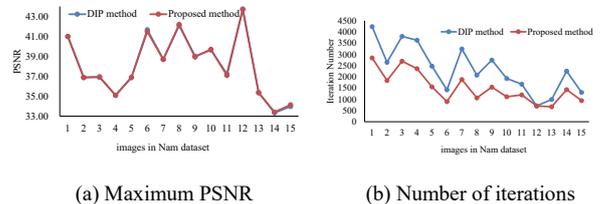

(a) Maximum PSNR  (b) Number of iterations
Fig. 6. Comparison of original DIP method and our proposed method after using transfer learning.

## IV. CONCLUSION

In this paper, we proposed a CNN-based blind denoising network for endoscopy images. For blind denoising, we optimize the DIP method by transfer learning to reduce the number of iteration. To determine the quality of reconstructed image, the blind image assessment network based on MobileNet is presented to estimate the scores of image quality. The experimental results show that our proposed method has a good noise suppression for the enhanced dark endoscopic images from the view of the visual quality. Moreover, our proposed method outperforms the state-of-the-arts on real-world noisy image datasets in terms of quantitative metrics.

TABLE I. THE QUANTITATIVE RESULTS ON THE NAM DATASET

| Camera | CBM3D | NC | NI | CC | Pro. |
|---|---|---|---|---|---|
| Canon 5D ISO = 3200 | 37.79 | 37.72 | 38.75 | 38.37 | **40.62** |
|  | 34.34 | 35.26 | 35.57 | 35.37 | **35.72** |
|  | 34.27 | 34.89 | 35.55 | 34.91 | **36.33** |
| Nikon D600 ISO = 3200 | 33.70 | 34.70 | **35.59** | 34.98 | 33.88 |
|  | 34.33 | 34.32 | 36.78 | 35.95 | **36.89** |
|  | 35.75 | 38.57 | 39.30 | 40.51 | **41.47** |
| Nikon D800 ISO= 1600 | 36.15 | 38.18 | 38.02 | 37.99 | **38.70** |
|  | 36.57 | 38.84 | 38.99 | 40.36 | **42.21** |
|  | 35.47 | 38.44 | 38.19 | 38.30 | **38.94** |
| Nikon D800 ISO = 3200 | 34.00 | 38.22 | 38.05 | 39.01 | **39.71** |
|  | 33.43 | 35.72 | 35.71 | 36.75 | **37.09** |
|  | 33.53 | 38.58 | 32.91 | 39.06 | **41.30** |
| Nikon D800 ISO = 6400 | 29.97 | 33.61 | 33.51 | 34.61 | **35.16** |
|  | 30.33 | 32.57 | 32.75 | 33.21 | **33.41** |
|  | 30.21 | 32.86 | 32.88 | 33.22 | **34.15** |
| Average | 33.99 | 36.17 | 35.33 | 36.88 | **37.56** |